\documentclass[10pt,twocolumn]{article}

\usepackage[margin=0.75in]{geometry}
\usepackage{amsmath}
\usepackage[authoryear,round]{natbib}
\usepackage{booktabs}
\usepackage{graphicx}
\usepackage{hyperref}
\usepackage{float}
\usepackage{longtable}

\title{Spatial Partner Markets: Effective Competition, Scale, and Opportunity}
\author{
{\Large Curren K. Mehta}\\
{\large Massachusetts Institute of Technology}
}
\date{\today}

\begin{document}

\twocolumn[
\begin{@twocolumnfalse}

\maketitle

\begin{center}
{\large\bfseries Abstract}
\end{center}

\noindent
Local partner markets differ not only in population size but in relative partner availability, the scale of opportunity, marriage timing, and population turnover. This paper develops a county-level framework for comparing these conditions across local partner markets. The empirical application focuses on unmarried women ages 25--34 and uses observed partner-age pairings to construct age-specific weights for the relevant unmarried male population. We model absolute opportunity separately, fitting a Michaelis--Menten diminishing-returns transformation to 192 scenarios that vary dating intensity, search horizon, fit, and first-date conversion. The resulting index also incorporates the share of women ages 35--39 who have never married and the inflow rate of adults ages 25--34. Among 320 counties with complete data, the District of Columbia ranks first and New York County second under the baseline specification, and rankings remain highly stable under alternative component weights. An endogenous mobility extension shows that favorable markets attract additional male participants, producing substantial adjustment while leaving the top composite rankings relatively stable over the parameter values considered. The framework is intended as a structural comparison of local partner-market conditions rather than a predictor of individual relationship outcomes.
\vspace{1em}
\end{@twocolumnfalse}
]

\section{Introduction}
Where a person lives changes the set of potential partners they can realistically meet. Yet local partner markets cannot be compared using population counts or simple sex ratios alone. A favorable ratio may occur in a small market, while a large city may contain both more potential partners and more competitors. Relevant competitors may also fall outside a narrowly defined age range, and the value of additional market size should diminish once the pool exceeds what an individual could realistically search.
\\
This paper develops a quantitative framework for comparing these local market conditions. The empirical application considers U.S. county-level partner markets from the perspective of men seeking women ages 25--34. The objective is not to predict whether a particular individual will form a relationship, but to compare structural differences in opportunity and competition across locations.
\\ \\
Prior research shows that partner markets are geographically local and that partner availability can affect marriage behavior \citep{lichter1991}. Partnerships are also strongly age-assortative, although meaningful age differences are common \citep{ausubel2022,geary2019}. A sex ratio that restricts both groups to the same age interval therefore ignores relevant competitors outside that interval, while treating all nearby ages equally overstates their importance. We address this by estimating age-specific competition weights from observed partner-age pairings in 2024 ACS microdata and applying them to county populations. This produces an effective competitor pool and a ratio of unmarried women ages 25--34 to effective unmarried male competitors.
\\
Competitive balance does not capture the absolute scale of opportunity. We therefore model the number of unmarried women ages 25--34 separately.
Because an individual cannot search an unlimited number of potential partners, market size is modeled with diminishing rather than constant returns. Sensitivity analysis across alternative assumptions about fit, first-date conversion, dating intensity, and search horizon leads us to select a Michaelis--Menten transformation for absolute opportunity.
\\ \\ 
Two additional measures capture how local markets persist and change over time: the share of women ages 35--39 who have never married and the rate at which adults ages 25--34 arrive from outside the county. These measures are combined with effective competition and absolute opportunity in a composite index. Under the baseline specification, the District of Columbia ranks first among the 320 counties with complete data and New York County second. The overall ranking is highly stable under alternative component weights. \\
Finally, we allow competitors to \textit{respond} to differences across markets. Favorable conditions can attract additional competitors, weakening the advantage that initially made a location attractive. In the equilibrium model, effective male competitors move toward markets with higher modeled dating-market attractiveness, which lowers their female-to-male ratios and creates a congestion effect. Increasing mobility responsiveness produces stronger counterfactual reallocation and can substantially narrow initial ratio differences, although effective-ratio dispersion need not decline monotonically. The highest-ranked markets remain relatively stable over the parameter values considered.
\\ \\
Taken together, the framework compares local partner markets through both relative competition and absolute opportunity while accounting for marriage timing, population turnover, and endogenous mobility.
\section{Constructing the Effective Dating-Market Ratio}
\subsection{Geographic Unit and Market Definition}
Counties and county-equivalent jurisdictions are used as the primary geographic unit. They provide a consistent unit across U.S. markets while avoiding the geographic breadth of many Metropolitan Statistical Areas (MSAs), which may include suburban and outlying counties connected to the urban core primarily through commuting ties \citep{censusmsa2026}. County boundaries are therefore treated as approximations of local partner markets, not as strict limits on partner search.
\\
The analysis includes all counties and county-equivalent jurisdictions for which the 2024 ACS 1-year B12002 table reports usable estimates. Throughout the analysis, \emph{unmarried} includes respondents who are never married, separated, widowed, or divorced; respondents currently married are excluded. After merging the additional component data, the final complete-case analytic sample contains 320 counties and county-equivalent jurisdictions with data available for all four index components. Percentile ranks reported for individual component measures are calculated
within the usable county sample for that component prior to the final
complete-case merge.

\subsection{Age Assortativity and Partner Age Differences}
Across current different-sex spouses and cohabiting partners in North America, men are on average approximately 2.2 years older than their female partners \citep{ausubel2022}. British data similarly show relatively small median age differences, while approximately one-third of opposite-sex partnerships differ by at least five years \citep{geary2019}. Men outside a
woman's own age group therefore remain relevant potential competitors, but should not necessarily be treated as equivalent to similarly aged men. \\
Studies of age preferences provide additional support for this distinction. Women generally report preferences for partners near their own age or somewhat older, while men's preferred age ranges shift with their own age \citep{kenrick1992,antfolk2017}. Because stated preferences do not map perfectly onto realized attraction or partnership behavior \citep{eastwick2025}, we estimate competition weights from observed partner-age pairings rather than imposing them from stated preferences.

\subsection{Empirical Estimation of Age Weights}
We define competitors as unmarried men who plausibly participate in the partner market for women ages 25--34, with their relevance varying by age. Age-specific weights are estimated using 2024 ACS 1-year microdata from IPUMS USA \citep{ruggles2025}. IPUMS's SPLOC linkage identifies co-resident spouses and unmarried partners, allowing each respondent's partner age to be
observed \citep{ipumssploc}.
\\
The sample is restricted to men ages 20--44 with an identified female spouse or partner, yielding 234,994 observed partnerships. All estimates use the IPUMS person weight (PERWT). The 20--44 age range is chosen to capture adjacent male age cohorts with meaningful observed overlap with women ages 25--34; it should not be interpreted as implying that competition is exactly zero outside this interval.
\\
For the five male age groups 20--24, 25--29, 30--34, 35--39, and
40--44, we estimate the share of partnered men whose female partner is ages 25--34. \\
Let $b$ denote the male age group, $p_i$ the person weight for man $i$, and $A_{p,i}$ the age of his linked female partner:
\begin{equation}
s_b =
\frac{
\sum_{i\in b} p_i \mathbf{1}\{25 \leq A_{p,i} \leq 34\}
}{
\sum_{i\in b} p_i
}.
\end{equation}
where $\mathbf{1}\{\cdot\}$ equals 1 when the condition holds and 0 otherwise. The numerator is the weighted number of partnered men in age group $b$ whose female partner is ages 25--34, while the denominator is the weighted number of all partnered men in that group.  Thus, $s_b$ measures each age group's overlap with the target female market independent of cohort size. Because these shares are estimated among men with observed partners, they are used here as relative age-overlap weights rather than as estimates of the probability that an unmarried man is actively dating.\\ 
We convert these shares into relative competition weights by normalizing to the age group with the greatest observed overlap:
\begin{equation}
\omega_b = \frac{s_b}{\max_j s_j}.
\end{equation}
The highest-overlap age group therefore receives a weight of 1, while the relative differences among age groups are preserved.
\begin{table}[H]
\centering
\caption{Empirical male age competition weights}
\label{tab:age_weights}
\small
\begin{tabular}{lcc}
\toprule
Male age & Partner share $s_b$ & Weight $\omega_b$ \\
\midrule
20--24 & 0.170 & 0.203 \\
25--29 & 0.796 & 0.949 \\
30--34 & 0.839 & 1.000 \\
35--39 & 0.345 & 0.411 \\
40--44 & 0.091 & 0.108 \\
\bottomrule
\end{tabular}
\end{table}
Men ages 30--34 have the greatest observed overlap with women ages 25--34 and therefore receive the reference weight of 1.000.
\subsection{Effective Dating-Market Ratio}
The estimated age weights are applied to county-level counts of unmarried men from the 2024 ACS. For county $c$, the effective male competitor population is:
\begin{equation}
\begin{aligned}
M_c^* ={}&
0.203M_{20\text{--}24,c}
+0.949M_{25\text{--}29,c}
+1.000M_{30\text{--}34,c} \\
&+0.411M_{35\text{--}39,c}
+0.108M_{40\text{--}44,c}.
\end{aligned}
\end{equation}
where $M_{a,c}$ is the number of unmarried men in age group $a$ and county $c$.
\\
The effective dating-market ratio is then:
\begin{equation}
R_c =
\frac{F_{25\text{--}34,c}}{M_c^*},
\end{equation}
where $F_{25\text{--}34,c}$ is the number of unmarried women ages 25--34. \\
For readability, results are reported as $100R_c$, representing unmarried women ages 25--34 per 100 effective male competitor units. Because the age weights are normalized to the highest-overlap age group, $M_c^*$ is measured in reference-group-equivalent units of competition rather than as a literal count of individual men. Accordingly, $100R_c$ should be interpreted as a comparative effective-competition ratio.
\begin{table}[H]
\centering
\caption{Selected county-level effective dating-market ratios}
\label{tab:selected_ratios}
\small
\begin{tabular}{llrr}
\toprule
Principal city & County & $100R_c$ & Pctl. \\
\midrule
Washington, DC & District of Columbia & 99.8 & 99.7 \\
Manhattan, NY & New York County, NY & 92.9 & 99.1 \\
Chicago & Cook County, IL & 77.1 & 86.1 \\
Dallas & Dallas County, TX & 71.9 & 70.1 \\
Miami & Miami-Dade County, FL & 67.3 & 45.4 \\
Phoenix & Maricopa County, AZ & 66.8 & 42.0 \\
San Jose & Santa Clara County, CA & 61.4 & 19.1 \\
\bottomrule
\end{tabular}
\end{table}
\section{Diminishing Returns to Market Size}
\subsection{The Linear Market-Size Benchmark}
The effective ratio captures competitive balance but not the absolute number of potential partners. Let $N_c^F$ denote the number of unmarried women ages 25--34 in county $c$. The simplest size-based benchmark is:
\begin{equation}
B_c = N_c^F.
\end{equation}
Under this specification, every additional potential partner contributes the same amount of opportunity regardless of market size.
\begin{table}[H]
\centering
\caption{Selected markets under a linear specification of unmarried women ages 25--34}
\label{tab:linear_market_size}
\small
\begin{tabular}{lrr}
\toprule
Principal city & Unmarried Women 25--34 & Linear index \\
\midrule
Los Angeles    & 538,379 & 100.0 \\
Chicago        & 288,387 & 53.6 \\
Manhattan, NY  & 147,322 & 27.4 \\
Miami          & 123,324 & 22.9 \\
Philadelphia   & 108,329 & 20.1 \\
Washington, DC & 65,684  & 12.2 \\
\bottomrule
\end{tabular}
\end{table}
\noindent
Under this specification, Los Angeles County receives approximately 3.7 times the size-based value of New York County and more than eight times
that of the District of Columbia. This follows directly from treating each additional potential partner as equally valuable.
\\
An individual, however, \textbf{cannot meet or evaluate an unlimited number of people}. Once a market is large enough to supply more plausible partners than an individual could realistically encounter, additional population should
have diminishing practical value.

\subsection{Translating Market Size Into Opportunities}
The raw number of unmarried women ages 25--34 \textit{overstates} the number of realistic dating opportunities available to any one individual. Only a fraction will satisfy basic compatibility criteria, and only a fraction of those plausible fits will result in a first date. The relevant question is therefore \textit{not} whether the entire female population is accessible for dating, but whether a sufficiently large raw market can still generate
a large number of actionable opportunities after these successive filters are applied.
\\
Let $p_{\mathrm{fit}}$ denote the probability that a randomly selected woman in the target population is a plausible dating candidate. Conditional on being a plausible fit, let $p_{\mathrm{match}}$ denote the effective probability that the candidate is encountered, reciprocal interest occurs, contact succeeds, and the pair converts to an in-person first date. The total number of dates an individual can pursue over time is modeled separately through $d$, $W$, and $T$. \\
The expected uncapped supply of first-date opportunities generated by county $c$ is
\begin{equation}
A_c = N_c^F p_{\mathrm{fit}} p_{\mathrm{match}}.
\end{equation}
Because $p_{\mathrm{match}}$ is conditional on being a plausible fit, the product follows directly from the probability chain rule. $A_c$ is an uncapped opportunity-supply measure rather than the expected number of dates actually realized by an individual; realized dating activity is constrained separately by the capacity $D$ defined below.
\\  \\
To determine when raw market size ceases to be the primary constraint, let $d$ denote new first dates per active dating week, $W$ active dating weeks per year, and $T$ the dating horizon in years. The corresponding first-date
capacity is
\begin{equation}
D = dWT.
\end{equation}
\noindent
Setting expected opportunity equal to dating capacity gives the corresponding critical market size
\begin{equation}
N^* =
\frac{D}{p_{\mathrm{fit}}p_{\mathrm{match}}}.
\end{equation}
Below $N^*$, additional potential partners may meaningfully expand the number of actionable opportunities. Above it, expected opportunity already exceeds modeled dating capacity, so additional market size should have progressively less practical value.
\\
Because these behavioral parameters cannot be observed precisely for a general individual, we evaluate them over a broad sensitivity range rather than selecting a single specification.

\subsection{Sensitivity Results and Observed Market Implications}
We first consider an illustrative reference case with
$p_{\mathrm{fit}}=5\%$ and $p_{\mathrm{match}}=5\%$. Under these
assumptions, 0.25\% of the raw population of unmarried women ages
25--34 is expected to generate an actionable first-date opportunity. We assume 40 active dating weeks per year and consider the three dating-capacity scenarios in Table~\ref{tab:dating_capacity}.
\begin{table}[H]
\centering
\caption{Selected dating-capacity scenarios}
\label{tab:dating_capacity}
\begin{tabular}{lrrrr}
\hline
Scenario & Dates/yr & Years & Total & $N^*$ \\
\hline
Moderate & 80 & 2 & 160 & 64,000 \\
High intensity & 120 & 2 & 240 & 96,000 \\
Large stress test & 160 & 3 & 480 & 192,000 \\
\hline
\end{tabular}
\end{table}
For example, under the moderate scenario, two first dates per active week over 40 active weeks produces 80 first dates per year, or 160 over two years. With $p_{\mathrm{fit}}=5\%$ and
$p_{\mathrm{match}}=5\%$, the corresponding critical market size is 
\begin{equation}
N^*
=
\frac{160}{0.05\times0.05}
=
64{,}000.
\end{equation}

Thus, a raw pool of 64,000 unmarried women ages 25--34 is expected to generate enough first-date opportunities to supply the modeled capacity. 
\\ 
Across the observed county data, 33 counties exceed this threshold, 16 exceed the 96,000-person high-intensity threshold, and four exceed the 192,000-person stress-test threshold.
Table~\ref{tab:expected_first_dates} applies the same reference assumptions to several large markets.
\begin{table}[H]
\centering
\caption{Expected first-date opportunities in selected markets}
\label{tab:expected_first_dates}
\begin{tabular}{@{}lrr@{}}
\hline
Principal city & Women 25--34 & Uncapped dates \\
\hline
Los Angeles    & 538,379 & 1,345.9 \\
Chicago        & 288,387 & 721.0 \\
Manhattan, NY  & 147,322 & 368.3 \\
Miami          & 123,324 & 308.3 \\
San Jose       & 84,886  & 212.2 \\
Washington, DC & 65,684  & 164.2 \\
\hline
\end{tabular}
\end{table}
\noindent
The contrast with the linear market-size benchmark is substantial.
Los Angeles County contains approximately 3.7 times as many unmarried women ages 25--34 as New York County. Under the reference assumptions, however, New York County generates
approximately 368 uncapped first-date opportunities, while Los Angeles
generates approximately 1,346. \\
Much of that additional supply exceeds even the intensive dating capacities considered above.
\\
These calculations do not identify a single market size beyond which additional population has no value. The threshold depends on fit, matching, dating intensity, and search horizon. They do show, however, that the \textit{marginal value} of additional population declines as markets grow. A proportional, unbounded market-size reward therefore \textit{overstates} the advantage of very large partner pools.

\subsection{Selecting a Diminishing Returns Function}
Rather than selecting one of the preceding scenarios as representative, we evaluate a broader grid of parameter combinations to construct the diminishing-returns function.
\\
The sensitivity grid uses the following assumptions:
\begin{itemize}
    \item Active dating weeks per year: 40
    \item New first dates per active week: 1, 2, 3, or 4
    \item Dating horizon: 1, 2, or 3 years
    \item $p_{\mathrm{fit}}$: 2.5\%, 5\%, 10\%, or 20\%
    \item $p_{\mathrm{match}}$: 2.5\%, 5\%, 10\%, or 20\%
\end{itemize}
\noindent
Because active dating weeks are held fixed, the four varying inputs generate
$4 \times 3 \times 4 \times 4 = 192$ distinct scenarios. These scenarios are treated as an equally weighted calibration grid rather than as an empirically estimated distribution of dating behavior. Each scenario $i$ produces a critical market size $N_i^*$ using the framework developed above. We translate this threshold into a market-size saturation profile.
\\
Let $N$ denote an arbitrary raw pool of unmarried women ages 25--34. For
scenario $i$, define
\begin{equation}
C_i(N)
=
\min\left\{
\frac{N}{N_i^*},
1
\right\}.
\end{equation}
The quantity $C_i(N)$ represents the fraction of modeled dating capacity supplied by a market of size $N$. It increases linearly until $N$ reaches the scenario-specific critical market size $N_i^*$, after which it remains equal to 1. For example, if $N_i^*=64{,}000$, a market of 32,000 women supplies one-half of the modeled capacity, so $C_i(32{,}000)=0.5$. \\
The 192 scenarios therefore produce 192 saturation profiles. Rather than select one as representative, we fit a single smooth diminishing-returns function to all profiles simultaneously.\\
Formally, let $g(N)$ denote a candidate smooth market-size function. We evaluate each candidate at $J=500$ equally weighted, evenly spaced female-pool sizes spanning the observed county range. Its fit is measured by the mean squared deviation from all 192
scenario profiles,
\begin{equation}
\operatorname{MSE}(g)
=
\frac{1}{192J}
\sum_{i=1}^{192}
\sum_{j=1}^{J}
\left[
g(N_j)-C_i(N_j)
\right]^2.
\end{equation}
For candidate functions with a free scale parameter, that parameter is
chosen to minimize this error.
\\ \\
We compare five candidate functions. Four impose positive but diminishing
marginal returns,
\[
g'(N)>0,
\qquad
g''(N)<0.
\]
\noindent
The linear specification, for which $g''(N)=0$, is retained as a benchmark.
Let $N_{\max}$ denote the largest observed female pool. The candidate
functions are:
\begin{align}
g_{\mathrm{linear}}(N)
&=
\frac{N}{N_{\max}}, \\[4pt]
g_{\mathrm{sqrt}}(N)
&=
\sqrt{\frac{N}{N_{\max}}}, \\[4pt]
g_{\mathrm{log}}(N;K)
&=
\frac{\log(1+N/K)}
{\log(1+N_{\max}/K)}, \\[4pt]
g_{\mathrm{MM}}(N;K)
&=
\frac{N}{N+K}, \\[4pt]
g_{\mathrm{exp}}(N;K)
&=
1-e^{-N/K}.
\end{align}
For the logarithmic, Michaelis--Menten, and exponential specifications,
the scale parameter $K$ is selected numerically to minimize this error. 
\begin{table}[H]
\centering
\caption{Fit of candidate diminishing-returns functions}
\label{tab:curve_fit}
\small
\begin{tabular}{lrrr}
\toprule
Function & Best $K$ & RMSE & MAE \\
\midrule
Michaelis--Menten & 10,798 & 0.151 & 0.087 \\
Exponential & 26,780 & 0.154 & 0.057 \\
Logarithmic & 9.98 & 0.162 & 0.104 \\
Square root & -- & 0.346 & 0.280 \\
Linear & -- & 0.514 & 0.436 \\
\bottomrule
\end{tabular}
\end{table} 
\noindent
The fit results narrow the choice primarily to the Michaelis--Menten and
exponential functions. Michaelis--Menten produces the lowest RMSE, while the
exponential function produces the lowest MAE. The difference between the two
is relatively small, so the fit statistics alone do not provide a strong reason
to select one over the other.
\\
We therefore compare the values produced by the two functions at several
representative market sizes: 
\begin{table}[H]
\centering
\caption{Comparison of diminishing-returns functions}
\label{tab:function_values}
\small
\setlength{\tabcolsep}{4pt}
\begin{tabular}{rccc}
\toprule
Female pool & Mean profile & M--M & Exponential \\
\midrule
5,000   & 0.279 & 0.316 & 0.170 \\
10,000  & 0.434 & 0.481 & 0.312 \\
25,000  & 0.662 & 0.698 & 0.607 \\
50,000  & 0.810 & 0.822 & 0.845 \\
100,000 & 0.915 & 0.903 & 0.976 \\
200,000 & 0.972 & 0.949 & 0.999 \\
\bottomrule
\end{tabular}
\end{table}
\noindent
The exponential function assigns less value than the mean profile at smaller market sizes and approaches saturation more quickly. At 10,000 women, for example, the mean profile is 0.434 compared with 0.312 under
the exponential function; at 100,000, the corresponding values are 0.915 and 0.976.
\\
Michaelis--Menten tracks the mean more evenly across the observed range
and also produces the lowest RMSE. Because the Michaelis--Menten and exponential specifications fit the calibration profiles similarly, the choice between them should be interpreted as a modeling specification rather than as uniquely identified by the data. We therefore define the county-level absolute-opportunity component as
\begin{equation}
O_c
\equiv
G(N_c^F)
=
\frac{N_c^F}{N_c^F+10{,}798}.
\end{equation}
The fitted scale parameter has a direct interpretation: $G(10{,}798)=0.5$. Additional population continues to increase opportunity beyond this point, but at a progressively slower rate.

\section{Late-Marriage Environment}
The preceding measures do not capture differences in local marriage timing.
In some markets, remaining unmarried into the late thirties is considerably
more common than in others, potentially sustaining a larger partner pool at
older ages.
\\
We measure this characteristic using the share of women ages 35--39 who
report never having married in the 2024 ACS. For county $c$, define
\begin{equation}
L_c =
\frac{N^{F,\mathrm{never}}_{35-39,c}}
     {N^{F}_{35-39,c}},
\end{equation}
where $N^{F,\mathrm{never}}_{35-39,c}$ is the number of never-married
women ages 35--39 and $N^{F}_{35-39,c}$ is the total female population
ages 35--39.
\\ 
Higher values of $L_c$ therefore indicate places in which a larger share
of women ages 35--39 has never married. We interpret this as a
cross-sectional indicator of the local late-marriage environment rather
than as a longitudinal measure of whether currently unmarried women ages
25--34 will remain unmarried or available in the same county. It is also
not evidence that residents are more likely to marry or form long-term
relationships, nor is it a direct estimate of age at first marriage or of
the probability that any individual will eventually marry.
\begin{table}[H]
\centering
\caption{Selected late-marriage environments}
\begin{tabular}{lrr}
\hline
Principal city & Never married 35--39 (\%) & Pctl. \\
\hline
Washington, DC & 52.6 & 98.5 \\
Manhattan, NY  & 48.9 & 97.2 \\
Chicago        & 39.4 & 87.0 \\
Dallas         & 34.8 & 75.9 \\
Miami          & 34.3 & 74.4 \\
Phoenix        & 28.2 & 46.0 \\
San Jose       & 22.2 & 22.8 \\
\hline
\end{tabular}
\end{table}
\section{Market Refresh}
The preceding measures describe the current dating market, but local partner
pools also differ in how quickly they are replenished by new residents. Using
the 2024 ACS 1-year table C07001, \emph{Geographic Mobility in the Past Year
by Age for Current Residence in the United States}, let
$I_{25\text{--}34,c}$ denote current residents ages 25--34 who lived outside
county $c$ one year earlier, and let $P_{25\text{--}34,c}$ denote the current
population ages 25--34. We define the county-level refresh rate as
\begin{equation}
F_c =
\frac{I_{25\text{--}34,c}}
     {P_{25\text{--}34,c}}.
\end{equation}
External inflow includes moves from another county in the same state, another
state, or abroad, while moves within the same county are excluded. The measure
therefore captures gross young-adult inflow rather than net population growth.
Because it is an all-sex population-flow measure, $F_c$ is interpreted as
market turnover and replenishment rather than as a direct measure of new
female partner supply.
\begin{table}[H]
\centering
\caption{Selected county-level market refresh rates}
\label{tab:refresh}
\begin{tabular}{lrr}
\hline
Principal city & Refresh rate (\%) & Pctl. \\
\hline
Manhattan, NY  & 14.5 & 82.5 \\
Dallas         & 12.5 & 69.7 \\
Washington, DC & 12.5 & 69.3 \\
San Jose       & 11.7 & 62.2 \\
Chicago        & 8.2  & 27.1 \\
Phoenix        & 8.0  & 25.0 \\
Miami          & 7.1  & 17.5 \\
\hline
\end{tabular}
\end{table}
\section{Composite Market Index}
The preceding sections have developed four complementary measures of local dating-market conditions: the balance between potential partners and effective competitors, the absolute scale of available opportunity, the prevalence of remaining never married into the late thirties, and the rate at which the young-adult population is replenished by new arrivals. Each
captures a different feature of the same underlying question: \textbf{how favorable is a local market for an individual searching for a long-term partner?}
\\ \\
We now bring these components together into a single composite index. The
goal is not to replace the underlying measures, each of which remains
individually interpretable, but to summarize their joint implications for
cross-market comparison. Constructing such an index requires two additional choices: placing the components on a common scale and specifying their relative importance.
\subsection{Standardization}
Let the four county-level components be denoted by $R_c$, $O_c$, $L_c$,
and $F_c$, where $R_c$ is the effective dating-market ratio, $O_c$ is
the diminished absolute-opportunity measure, $L_c$ is the late-marriage
measure, and $F_c$ is the market refresh rate. All four are defined such
that larger values represent more favorable market conditions.
\\
Because the components are measured on different scales, we standardize
each across the county sample. For component $k$ in county $c$,
\begin{equation}
Z_{kc}
=
\frac{X_{kc}-\bar{X}_k}{s_k},
\end{equation}
where $\bar{X}_k$ and $s_k$ are the sample mean and standard deviation of
component $k$. A one-unit change in a standardized component therefore
represents a one-standard-deviation difference across counties.
\\ 
For absolute opportunity, standardization is applied after the
Michaelis--Menten transformation developed in Section~3. Because
standardization is a positive affine transformation, the ordering and
diminishing-return shape of $O_c$ are preserved, but the remaining
cross-county variation is rescaled to unit standard deviation. Accordingly,
the 35\% opportunity weight should be interpreted as a weight on standardized
cross-county variation after diminishing returns, rather than as a direct
percentage contribution of raw market size.
\subsection{Baseline Weights}
The four components do not represent equally direct measures of dating
opportunity. We therefore assign weights according to a simple
hierarchical structure rather than weighting all four equally.
\\
The effective ratio and absolute-opportunity measure describe the
current structure of the partner market and jointly receive 80\% of the
baseline index. Within this group, slightly greater weight is assigned
to the effective ratio because it directly measures the balance between
the target population and the empirically weighted competitor pool.
Absolute market size remains important, but its marginal value declines
as the pool becomes sufficiently large.
\\
The remaining 20\% is assigned to characteristics that describe the
market over a longer search horizon. The late-marriage measure receives 15\%. Its positive contribution is an explicit modeling assumption: a higher prevalence of never-married women ages 35--39 is treated as evidence of a local environment in which
unmarried status remains more common into later ages, thereby extending
the potential availability of unmarried partners over a longer search
horizon. This interpretation does not imply that such markets have higher
rates of relationship or marriage formation. Refresh receives 5\%. Because the measure captures all-sex inflow and does not distinguish relationship status, its positive contribution to the index is an explicit modeling assumption that greater young-adult turnover expands the rate at which the local social and partner market is replenished. It should \textit{not} be interpreted as a direct measure of new female partner supply and is
therefore assigned the smallest baseline weight.
\\
This yields the \textbf{baseline composite index:}
\begin{equation}
I_c =
0.45Z(R_c)
+
0.35Z(O_c)
+
0.15Z(L_c)
+
0.05Z(F_c).
\end{equation}
The weights sum to one and reflect the intended conceptual hierarchy:
current competition and opportunity dominate the index, while
late-marriage prevalence and population refresh modify the assessment
of otherwise similar markets.
\\
For presentation, the composite values are rescaled to a 0--100 range:
\begin{equation}
S_c =
100
\frac{
I_c-\min_j I_j
}{
\max_j I_j-\min_j I_j
}.
\end{equation}
This final rescaling affects only the presentation of the index and not
the ordering of markets.

\subsection{Baseline Composite Results}
Applying the baseline weights produces a composite index for the 320
counties with complete data on all four measures.
\\ 
Table~\ref{tab:baseline_composite} reports results for the selected markets
used throughout the analysis. The District of Columbia ranks first in the
county sample, followed by New York County. Chicago ranks 18th, Dallas
31st, Miami-Dade 64th, Maricopa 68th, and Santa Clara 139th.
\begin{table}[H]
\centering
\caption{Baseline composite results for selected markets}
\label{tab:baseline_composite}
\scriptsize
\resizebox{\columnwidth}{!}{%
\begin{tabular}{@{}lrrrrrr@{}}
\hline
City & Rank & Index & Ratio & Opp. & Late (\%) & Refresh (\%) \\
\hline
Washington, DC & 1   & 100.0 & 99.8 & 0.859 & 52.6 & 12.5 \\
Manhattan, NY  & 2   & 95.1  & 92.9 & 0.932 & 48.9 & 14.5 \\
Chicago        & 18  & 74.4  & 77.1 & 0.964 & 39.4 & 8.2  \\
Dallas         & 31  & 66.9  & 71.9 & 0.928 & 34.8 & 12.5 \\
Miami          & 64  & 59.9  & 67.3 & 0.919 & 34.3 & 7.1  \\
Phoenix        & 68  & 58.7  & 66.8 & 0.949 & 28.2 & 8.0  \\
San Jose       & 139 & 49.1  & 61.4 & 0.887 & 22.2 & 11.7 \\
\hline
\end{tabular}}
\end{table}
\noindent
Here, Ratio denotes unmarried women ages 25--34 per 100 effective male
competitors, Opp.\ denotes the Michaelis--Menten-transformed absolute
opportunity measure, Late denotes the percentage of women ages 35--39
who have never married, and Refresh denotes the annual external-inflow
share of the population ages 25--34.
\subsection{Weight Sensitivity}
Because the baseline weights are conceptually specified rather than
empirically estimated, we test six alternatives. In
(ratio, opportunity, late-marriage, refresh) order, the specifications are
Ratio+ = $(0.50,0.30,0.15,0.05)$, Opp.+ = $(0.40,0.40,0.15,0.05)$,
Late+ = $(0.45,0.30,0.20,0.05)$, Refresh+ = $(0.45,0.30,0.15,0.10)$,
No ref. = $(0.45,0.40,0.15,0)$, and Core only =
$(0.55,0.45,0,0)$.
\\
Across the five nearby alternative specifications, the correlation
between county ranks and the baseline ranking ranges from 0.995 to 0.997.
Even the substantially different core-measures-only specification produces
a rank correlation of 0.981. The overall ordering of counties is therefore
highly stable under the alternative weights considered. Table~\ref{tab:weight_sensitivity} shows how these changes affect the selected markets used throughout the analysis.
\begin{table}[H]
\centering
\caption{Overall county rank under alternative weighting specifications}
\label{tab:weight_sensitivity}
\scriptsize
\resizebox{\columnwidth}{!}{
\begin{tabular}{lrrrrrrr}
\hline
City & Base & Ratio+ & Opp.+ & Late+ & Refresh+ & No ref. & Core only \\
\hline
Washington, DC & 1   & 1   & 1   & 1   & 1   & 1   & 1   \\
Manhattan, NY  & 2   & 2   & 2   & 2   & 2   & 2   & 2   \\
Chicago        & 18  & 22  & 15  & 22  & 21  & 13  & 13  \\
Dallas         & 31  & 37  & 30  & 35  & 37  & 30  & 34  \\
Miami          & 64  & 77  & 55  & 68  & 79  & 52  & 64  \\
Phoenix        & 68  & 90  & 62  & 89  & 93  & 61  & 57  \\
San Jose       & 139 & 166 & 119 & 166 & 155 & 130 & 124 \\
\hline
\end{tabular}
}
\end{table}
\noindent
Base denotes the 45/35/15/5 baseline specification; the remaining columns
use the alternative weight vectors defined above. Rank 1 denotes the
highest-ranked county among the 320 counties with complete data.
\\ 
Stability is particularly strong at the top of the distribution.
The District of Columbia ranks first and New York County second
under every specification considered. Chicago remains between 13th and
22nd, while Dallas remains between 30th and 37th. Greater variation
appears farther down the ranking: Miami-Dade ranges from 52nd to 79th,
Maricopa from 57th to 93rd, and Santa Clara from 119th to 166th.
\\
The broad ordering is therefore robust to the alternative weights, although
weighting choices matter more among middle- and lower-ranked markets.

\section{Endogenous Mobility and Spatial Equilibrium}
The composite index above treats county dating markets as static. If
individuals migrate in response to differences in dating-market opportunity,
attractive markets may draw additional competitors and thereby weaken their
initial advantage. We therefore allow the effective male competitor
population to respond endogenously to local dating-market attractiveness.

\subsection{Mobility Model}
We model effective male competitors as a population distributed across
counties. Rather than simulate individuals separately, we use a mean-field
representation in which each county's effective competitor mass evolves with
aggregate location choices. \\
Let $M_c^*(t)$ denote county $c$'s effective male competitor population
at iteration $t$. Holding the female partner pool fixed, the effective
dating-market ratio becomes:
\begin{equation}
R_c(t)
=
\frac{N_c^F}{M_c^*(t)},
\end{equation}
where $N_c^F$ is the number of unmarried women ages 25--34. For readability, equilibrium results are again reported as $100R_c(t)$.\\
Absolute opportunity, late-marriage environment, and refresh are held fixed,
so the standardized dating-market attractiveness index at iteration $t$ is
\begin{equation}
\begin{aligned}
U_c(t) ={}& 0.45Z_R(R_c(t)) + 0.35Z_O(O_c) \\
          &+ 0.15Z_L(L_c) + 0.05Z_F(F_c).
\end{aligned}
\end{equation}
The standardization parameters for the ratio are fixed at their baseline
values so that the scale of the index does not change as the population
redistributes. Because the component weights are conceptually specified
rather than estimated from observed location choices, $U_c(t)$ should be
interpreted as a stylized standardized attractiveness index rather than as
empirically estimated cardinal utility.
\\
The simulation is initialized at the observed effective competitor
distribution, $M_c^*(0)=M_{c,\mathrm{obs}}^*$. Let
$M_{\mathrm{tot}}^*=\sum_c M_c^*(0)$ denote the fixed total effective
competitor mass across the analytic sample. Location choice is modeled
probabilistically, with
\begin{equation}
b_c=\frac{M_c^*(0)}{M_{\mathrm{tot}}^*}
\end{equation}
denoting county $c$'s initial share of effective competitor mass. This baseline distribution is treated as the observed reference distribution of effective competitors across counties. It summarizes the combined influence of all historical determinants of residential location, which may include employment, housing, family, amenities, moving costs, and existing dating-market conditions. The mobility exercise therefore should be interpreted as a stylized counterfactual reallocation relative to the observed distribution as increasing weight is placed on the modeled dating-market attractiveness index, rather than as an identified decomposition of dating and non-dating location preferences.\\ 
The probability that competitor mass is allocated to county
$c$ is
\begin{equation}
P_c(t)
=
\frac{
b_c\exp\{\beta U_c(t)\}
}{
\sum_j b_j\exp\{\beta U_j(t)\}
},
\end{equation}
where $\beta$ controls the responsiveness of location choice to
dating-market attractiveness. When $\beta=0$, the observed reference
distribution is recovered. Larger values of $\beta$ place progressively
greater weight on differences in the modeled dating-market attractiveness
index relative to that reference distribution, generating stronger
counterfactual reallocation.
\\
Competitor populations adjust gradually toward the distribution implied
by these probabilities:
\begin{equation}
M_c^*(t+1)
=
(1-\mu)M_c^*(t)
+
\mu M_{\mathrm{tot}}^*P_c(t),
\end{equation}
where $\mu$ is an adjustment parameter and $M_{\mathrm{tot}}^*$ is the
fixed total effective competitor mass across the analytic sample. The
baseline simulation uses $\beta=1$ and $\mu=0.10$. Iterations are
computational adjustment steps and are not interpreted as calendar years.
\\
This structure creates an endogenous congestion mechanism. A market with
high initial modeled attractiveness attracts additional competitors, which lowers its
female-to-effective-male ratio and therefore reduces its subsequent attractiveness. Conversely, markets with relatively low attractiveness lose competitor
mass, improving their effective ratio. The system is iterated until the maximum absolute change in any county's share of total effective competitor mass between successive iterations falls below $10^{-10}$.

\subsection{Baseline Equilibrium Results}
The equilibrium simulation produces substantial redistribution of
effective competitor mass while sharply compressing differences in the
effective dating-market ratio. Table~\ref{tab:equilibrium_results} reports results for the selected markets used throughout the analysis.
\begin{table}[H]
\centering
\caption{Selected markets before and after endogenous competitor adjustment}
\label{tab:equilibrium_results}
\scriptsize
\resizebox{\columnwidth}{!}{%
\begin{tabular}{@{}lrrrrr@{}}
\hline
City & Base rank & Eq. rank & Base ratio & Eq. ratio & $\Delta M^*$ (\%) \\
\hline
Washington, DC & 1   & 1   & 99.8 & 67.7 & +47.4 \\
Manhattan, NY  & 2   & 2   & 92.9 & 64.8 & +43.3 \\
Chicago        & 18  & 17  & 77.1 & 64.9 & +18.8 \\
Dallas         & 31  & 31  & 71.9 & 65.2 & +10.2 \\
Miami          & 64  & 64  & 67.3 & 65.8 & +2.3  \\
Phoenix        & 68  & 68  & 66.8 & 66.2 & +1.0  \\
San Jose       & 139 & 154 & 61.4 & 67.6 & -9.2  \\
\hline
\end{tabular}
}
\end{table}
\noindent
Here, Base ratio and Eq.\ ratio denote unmarried women ages 25--34 per
100 effective male competitors before and after adjustment, respectively,
while $\Delta M^*$ denotes the percentage change in effective competitor
mass.
\\
The adjustment follows the model's congestion mechanism. Washington, DC
and New York County attract the largest competitor inflows among the selected
markets, increasing effective male competitor mass by 47.4\% and 43.3\%
and reducing their effective ratios from 99.8 to 67.7 and 92.9 to 64.8,
respectively. Chicago gains 18.8\% in competitor mass, reducing its ratio
from 77.1 to 64.9. 
\\
Adjustment is smaller in middle-ranked markets. Dallas gains 10.2\% in
effective competitor mass, while Miami and Phoenix gain only 2.3\% and
1.0\%. Santa Clara moves in the opposite direction, losing 9.2\% and
improving its ratio from 61.4 to 67.6.
\\
Among these selected markets, effective ratios compress from an initial
range of 61.4--99.8 to approximately 64.8--67.7. The model therefore
produces a partially self-correcting response: favorable conditions attract
competition, while unfavorable conditions encourage competitor outflow.
\\
The adjustment does not eliminate all cross-market differences under the baseline mobility specification. The District of Columbia remains first and New York County second at equilibrium. However, the fact that absolute opportunity, late-marriage prevalence, and market refresh are held fixed does not imply that their contribution to overall market attractiveness is insulated from competitive adjustment: additional competitor inflows can offset part of the advantage created by those characteristics through a less favorable effective ratio. The model therefore distinguishes fixed structural characteristics from the endogenous competitive response they induce.
\subsection{Sensitivity to Mobility Responsiveness}
Because $\beta$ is not empirically estimated, we evaluate
$\beta \in \{0, 0.25, 0.5, 1, 2\}$. At $\beta=0$, the model applies no
additional reallocation based on the dating-market attractiveness index,
and the observed reference distribution is recovered. Larger values of
$\beta$ place progressively greater weight on differences in modeled
dating-market attractiveness and generate stronger counterfactual
reallocation across counties.
\\
As $\beta$ increases, favorable markets attract more competitors and their
effective ratios fall. Washington, DC declines from 99.8 at $\beta=0$ to
78.6 at $\beta=0.25$, 67.7 at $\beta=1$, and 64.6 at $\beta=2$, while
competitor mass rises by 54.7\%. New York County similarly falls from 92.9
to 61.9 as competitor mass rises by 50.0\%.
\\
Less favorable markets move in the opposite direction. Santa Clara County's
effective ratio rises from 61.4 at $\beta=0$ to 68.6 at $\beta=2$ as
competitor mass falls, while Miami-Dade and Maricopa Counties change little.
\begin{table}[H]
\centering
\caption{Equilibrium effective ratio under alternative mobility responsiveness}
\scriptsize
\begin{tabular}{lrrrrr}
\hline
City & $\beta=0$ & $\beta=0.25$ & $\beta=0.5$ & $\beta=1$ & $\beta=2$ \\
\hline
Washington, DC & 99.8 & 78.6 & 72.5 & 67.7 & 64.6 \\
Manhattan, NY  & 92.9 & 74.6 & 69.1 & 64.8 & 61.9 \\
Chicago        & 77.1 & 69.4 & 66.9 & 64.9 & 63.6 \\
Dallas         & 71.9 & 67.6 & 66.2 & 65.2 & 64.5 \\
Miami          & 67.3 & 66.2 & 65.9 & 65.8 & 65.8 \\
Phoenix        & 66.8 & 66.2 & 66.1 & 66.2 & 66.2 \\
San Jose       & 61.4 & 64.8 & 66.2 & 67.6 & 68.6 \\
\hline
\end{tabular}
\end{table}
\noindent
Relative to the baseline distribution, endogenous mobility substantially
narrows differences in effective ratios. The overall composite ranking is
nevertheless relatively stable across the values of $\beta$ considered:
the District of Columbia ranks first and New York County second throughout,
while Chicago moves only modestly and Dallas, Miami-Dade, and Maricopa
retain their baseline ranks.
\\
The ratio adjustment itself is not monotonic in $\beta$. At stronger mobility
responsiveness, some effective-ratio differences reverse or widen as
competitor inflows increasingly offset advantages arising from the fixed
components of market attractiveness. The model therefore predicts convergence
in overall modeled attractiveness more directly than convergence in the
effective-ratio component alone.
\section{Assumptions, Limitations, and Future Work}
The framework relies on several simplifying assumptions. Counties are used
as approximations of local dating markets, although actual search areas may
cross county boundaries or be much smaller than a county. The analysis is
also based primarily on 2024 ACS data and therefore represents a
cross-sectional snapshot rather than a permanent characterization of each
market.\\
The ACS marital-status data also \textit{do not} identify whether legally unmarried
respondents are currently partnered, actively seeking a partner, or
participating in the modeled opposite-sex dating market. The unmarried
population should therefore be interpreted as a proxy for potential
availability rather than a direct, empirical count of active daters. 
\\
Several model components are approximations rather than directly observed
behavioral parameters. Age-specific competition weights are estimated from
existing partnerships and applied uniformly across counties, so geographic
variation in age-pairing patterns is not modeled. The absolute-opportunity
transformation relies on a range of assumed fit and first-date conversion
probabilities. \\
The 45/35/15/5 composite weights are conceptually specified rather than
empirically estimated, although the sensitivity analysis shows that the
broad ranking is not strongly dependent on their exact values. Similarly,
the mobility parameter $\beta$ is used to study how equilibrium changes
under different levels of responsiveness rather than to forecast actual
migration. 
\\ The ACS and IPUMS inputs are survey estimates, but the framework treats them
as point estimates when constructing competition weights, component values,
and county rankings. Sampling uncertainty is therefore not propagated through
the final index, so small differences between closely ranked counties should
not be interpreted as statistically precise.
\\
The composite index is not calibrated or validated against an observed
relationship-formation outcome. It should therefore be interpreted as a
structural measure of local dating-market conditions rather than as a
predictive model of individual dating or relationship success.\\
The equilibrium model is intentionally simple. It redistributes aggregate
effective competitor mass rather than age-specific male populations, thereby
treating competitor-equivalent units as homogeneous and abstracting from
changes in the age composition of movers across destinations. Effective male
competitors can relocate in response to dating-market attractiveness, while
the female population and other market characteristics remain fixed. Real
migration decisions also depend on wages, housing costs, family ties,
amenities, moving costs, and individual preferences, none of which are
modeled separately here.
\\
Future work could relax these assumptions using longitudinal or
individual-level data. Dating-platform or survey data could provide direct
estimates of matching behavior and market responsiveness, while repeated
ACS observations could test whether changes in local partner-market
conditions predict subsequent migration. A richer spatial model could also
allow both sides of the market to move and weight potential partners by
travel time rather than county boundaries.

\section{Conclusion}
This paper develops a quantitative framework for comparing local partner
markets by combining competitive balance, market size, marriage timing, and
population turnover. Rather than treating nearby male age groups as equally relevant competitors, the analysis estimates age-specific competition weights from observed
partner-age pairings. Absolute market size is then
adjusted for diminishing returns, while late-marriage prevalence and population
refresh capture additional features of the local environment.
\\
The resulting composite index produces substantial differences across U.S.
counties. Under the baseline specification, the District of Columbia and New York
County rank first and second, respectively, while Cook County ranks 18th. Sensitivity analysis shows that the broad ordering is highly stable under alternative component weights, although the precise placement of middle- and lower-ranked markets is more dependent on the weighting scheme.
\\
The equilibrium extension shows that these market advantages are not entirely
static. When effective male competitors are allowed to respond to local
dating-market attractiveness, favorable markets attract additional competition
and their initial effective-ratio advantages narrow. As mobility responsiveness
increases, competitor reallocation increasingly offsets differences in overall
modeled attractiveness, although effective-ratio differences need not shrink
monotonically. The top rankings remain relatively stable across the parameter
values considered.
\\
The framework therefore suggests that local partner markets are shaped by both
current opportunity and endogenous adjustment. Geographic advantages can be
meaningful, but they are neither one-dimensional nor fixed. The model provides
a transparent way to measure those differences while also showing how migration
can partially arbitrage them over time.

\bibliographystyle{apalike}
\bibliography{references}

@article{lichter1991,
  author  = {Lichter, Daniel T. and LeClere, Felicia B. and McLaughlin, Diane K.},
  title   = {Local Marriage Markets and the Marital Behavior of Black and White Women},
  journal = {American Journal of Sociology},
  year    = {1991},
  volume  = {96},
  number  = {4},
  pages   = {843--867},
  doi     = {10.1086/229610}
}

@misc{censusmsa2026,
  author       = {{U.S. Census Bureau}},
  title        = {Metropolitan and Micropolitan Statistical Area Glossary},
  year         = {2026},
  howpublished = {\url{https://www.census.gov/programs-surveys/metro-micro/about/glossary.html}},
}

@article{ausubel2022,
  author  = {Ausubel, Jacob and Kramer, Stephanie and Shi, Anne Fengyan and Hackett, Conrad},
  title   = {Measuring Age Differences among Different-Sex Couples: Across Religions and 130 Countries, Men Are Older than Their Female Partners},
  journal = {Population Studies},
  year    = {2022},
  volume  = {76},
  number  = {3},
  pages   = {465--476},
  doi     = {10.1080/00324728.2022.2094452}
}

@article{geary2019,
  author  = {Geary, Rebecca S. and Copas, Andrew J. and Sonnenberg, Pam and Tanton, Clare and King, Eleanor and Jones, Kyle G. and Trifonova, Viktoriya and Johnson, Anne M. and Mercer, Catherine H.},
  title   = {Sexual Mixing in Opposite-Sex Partnerships in Britain and Its Implications for STI Risk: Findings from the Third National Survey of Sexual Attitudes and Lifestyles (Natsal-3)},
  journal = {International Journal of Epidemiology},
  year    = {2019},
  volume  = {48},
  number  = {1},
  pages   = {228--242},
  doi     = {10.1093/ije/dyy237}
}

@article{kenrick1992,
  author  = {Kenrick, Douglas T. and Keefe, Richard C.},
  title   = {Age Preferences in Mates Reflect Sex Differences in Human Reproductive Strategies},
  journal = {Behavioral and Brain Sciences},
  year    = {1992},
  volume  = {15},
  number  = {1},
  pages   = {75--91},
  doi     = {10.1017/S0140525X00067595}
}

@article{antfolk2017,
  author  = {Antfolk, Jan},
  title   = {Age Limits: Men's and Women's Youngest and Oldest Considered and Actual Sex Partners},
  journal = {Evolutionary Psychology},
  year    = {2017},
  volume  = {15},
  number  = {1},
  doi     = {10.1177/1474704917690401}
}

@article{eastwick2025,
  author  = {Eastwick, Paul W. and Finkel, Eli J. and Meza, Eva M. and Ammerman, Kellie},
  title   = {No Gender Differences in Attraction to Young Partners: A Study of 4500 Blind Dates},
  journal = {Proceedings of the National Academy of Sciences},
  year    = {2025},
  volume  = {122},
  number  = {5},
  pages   = {e2416984122},
  doi     = {10.1073/pnas.2416984122}
}

@misc{ruggles2025,
  author    = {Ruggles, Steven and Flood, Sarah and Sobek, Matthew and
               Backman, Daniel and Cooper, Grace and Rivera Drew, Julia A.
               and Richards, Stephanie and Rogers, Renae and Schroeder,
               Jonathan and Williams, Kari C. W.},
  title     = {IPUMS USA: Version 16.0},
  year      = {2025},
  publisher = {IPUMS},
  address   = {Minneapolis, MN},
  doi       = {10.18128/D010.V16.0}
}

@misc{ipumssploc,
  author       = {{IPUMS USA}},
  title        = {SPLOC: Spouse's Location in Household},
  year         = {2026},
  howpublished = {\url{https://usa.ipums.org/usa-action/variables/sploc}},
  note         = {IPUMS USA documentation, accessed September 28, 2026}
}

\onecolumn

\section*{Appendix: Selected Baseline Rankings}

For reference, Table~\ref{tab:full_baseline_ranking} reports the top 90
counties and county-equivalent jurisdictions under the baseline 45/35/15/5
weighting specification. The reported score is the composite index rescaled
to the 0--100 presentation scale defined in Equation~(22); this rescaling
does not affect county ordering.

\begin{longtable}{r p{2.7cm} p{5.4cm} r}
\caption{Selected county ranking under the 45/35/15/5 baseline}\label{tab:full_baseline_ranking}\\
\hline
Rank & Principal city/area & County & Score \\
\hline
\endfirsthead
\multicolumn{4}{c}{{\tablename\ \thetable{} -- continued}}\\
\hline
Rank & Principal city/area & County & Score \\
\hline
\endhead
\hline
\multicolumn{4}{r}{Continued on next page}\\
\endfoot
\hline
\endlastfoot
1 & Washington, DC & District of Columbia, District of Columbia & 100.0 \\
2 & Manhattan, NY & New York County, New York & 95.1 \\
3 & Philadelphia & Philadelphia County, Pennsylvania & 87.4 \\
4 & Baltimore & Baltimore city, Maryland & 85.7 \\
5 & Brooklyn, NY & Kings County, New York & 84.8 \\
6 & Jackson & Hinds County, Mississippi & 84.8 \\
7 & Montgomery & Montgomery County, Alabama & 80.7 \\
8 & Boston & Suffolk County, Massachusetts & 78.2 \\
9 & Nashville, TN & Davidson County, Tennessee & 77.6 \\
10 & Stonecrest & DeKalb County, Georgia & 76.7 \\
11 & Columbia & Richland County, South Carolina & 76.4 \\
12 & Milwaukee & Milwaukee County, Wisconsin & 76.3 \\
13 & Towson, MD & Baltimore County, Maryland & 76.3 \\
14 & Bronx, NY & Bronx County, New York & 75.6 \\
15 & Charlotte & Mecklenburg County, North Carolina & 75.5 \\
16 & Atlanta & Fulton County, Georgia & 75.4 \\
17 & Memphis & Shelby County, Tennessee & 75.1 \\
18 & Chicago & Cook County, Illinois & 74.4 \\
19 & St. Louis & St. Louis city, Missouri & 74.2 \\
20 & Savannah & Chatham County, Georgia & 73.9 \\
21 & Detroit & Wayne County, Michigan & 73.9 \\
22 & Cleveland & Cuyahoga County, Ohio & 72.9 \\
23 & New Orleans & Orleans Parish, Louisiana & 71.8 \\
24 & Indianapolis, IN & Marion County, Indiana & 71.7 \\
25 & Portland & Multnomah County, Oregon & 70.7 \\
26 & Los Angeles & Los Angeles County, California & 70.6 \\
27 & Camden & Camden County, New Jersey & 70.3 \\
28 & Mobile & Mobile County, Alabama & 70.0 \\
29 & Denver & Denver County, Colorado & 68.4 \\
30 & Queens, NY & Queens County, New York & 68.0 \\
31 & Dallas & Dallas County, Texas & 66.9 \\
32 & San Francisco & San Francisco County, California & 66.8 \\
33 & Jacksonville & Duval County, Florida & 66.6 \\
34 & Baton Rouge & East Baton Rouge Parish, Louisiana & 65.7 \\
35 & Arlington & Arlington County, Virginia & 65.6 \\
36 & Columbus & Franklin County, Ohio & 65.6 \\
37 & Florissant & St. Louis County, Missouri & 65.3 \\
38 & Bowie & Prince George's County, Maryland & 65.2 \\
39 & Houston & Harris County, Texas & 64.3 \\
40 & Augusta, GA & Richmond County, Georgia & 64.1 \\
41 & St. Paul & Ramsey County, Minnesota & 64.0 \\
42 & Portland & Cumberland County, Maine & 63.8 \\
43 & Birmingham & Jefferson County, Alabama & 63.8 \\
44 & Durham & Durham County, North Carolina & 63.7 \\
45 & Oakland & Alameda County, California & 63.3 \\
46 & Tampa & Hillsborough County, Florida & 63.1 \\
47 & Las Vegas & Clark County, Nevada & 63.0 \\
48 & Louisville, KY & Jefferson County, Kentucky & 63.0 \\
49 & Fort Lauderdale & Broward County, Florida & 62.6 \\
50 & Jersey City & Hudson County, New Jersey & 62.4 \\
51 & Cambridge & Middlesex County, Massachusetts & 62.3 \\
52 & Orlando & Orange County, Florida & 61.9 \\
53 & Pittsburgh & Allegheny County, Pennsylvania & 61.7 \\
54 & Kansas City & Jackson County, Missouri & 61.6 \\
55 & Greensboro & Guilford County, North Carolina & 61.4 \\
56 & Anaheim & Orange County, California & 61.3 \\
57 & Metairie & Jefferson Parish, Louisiana & 61.2 \\
58 & Yonkers & Westchester County, New York & 61.0 \\
59 & San Bernardino & San Bernardino County, California & 60.9 \\
60 & Minneapolis & Hennepin County, Minnesota & 60.4 \\
61 & Providence & Providence County, Rhode Island & 60.4 \\
62 & Raleigh & Wake County, North Carolina & 60.2 \\
63 & West Palm Beach & Palm Beach County, Florida & 60.0 \\
64 & Miami & Miami-Dade County, Florida & 59.9 \\
65 & Fort Worth & Tarrant County, Texas & 59.5 \\
66 & Quincy & Norfolk County, Massachusetts & 59.3 \\
67 & Cincinnati & Hamilton County, Ohio & 58.8 \\
68 & Phoenix & Maricopa County, Arizona & 58.7 \\
69 & Madera & Madera County, California & 58.6 \\
70 & Austin & Travis County, Texas & 58.6 \\
71 & Buffalo & Erie County, New York & 58.5 \\
72 & Mableton & Cobb County, Georgia & 58.5 \\
73 & Rochester & Monroe County, New York & 58.4 \\
74 & Seattle & King County, Washington & 58.4 \\
75 & Springfield & Hampden County, Massachusetts & 58.3 \\
76 & New Haven & South Central Connecticut Planning Region, Connecticut & 58.3 \\
77 & Fresno & Fresno County, California & 58.2 \\
78 & Port St. Lucie & St. Lucie County, Florida & 58.1 \\
79 & Charleston & Charleston County, South Carolina & 58.1 \\
80 & Albuquerque & Bernalillo County, New Mexico & 58.0 \\
81 & Newark & Essex County, New Jersey & 58.0 \\
82 & San Mateo & San Mateo County, California & 57.8 \\
83 & St. Petersburg & Pinellas County, Florida & 57.6 \\
84 & Stamford & Western Connecticut Planning Region, Connecticut & 57.6 \\
85 & San Diego & San Diego County, California & 57.6 \\
86 & Sacramento & Sacramento County, California & 57.4 \\
87 & Dayton & Montgomery County, Ohio & 57.3 \\
88 & Riverside & Riverside County, California & 57.1 \\
89 & Albany & Albany County, New York & 57.1 \\
90 & Winston-Salem & Forsyth County, North Carolina & 57.0 \\
\end{longtable}

\end{document}